\documentclass[a4paper,10pt]{article}
\usepackage{amsmath,amsfonts,amsthm,amssymb,graphicx,hyperref}
\usepackage{xy,rotating}
\usepackage{pstricks}
\usepackage{mathrsfs}
\usepackage{palatino, url, multicol}
\usepackage{algorithm,algorithmic}
\usepackage{float}

\theoremstyle{plain}
\newtheorem{thm}{Theorem}

\newtheorem{lem}{Lemma}

\theoremstyle{definition}
\newtheorem{defn}{Definition}

\newcommand{\bm}[1]{\boldsymbol{#1}}

\begin{document}
\title{Approximated segmentation considering technical and dosimetric constraints in intensity-modulated radiation therapy with electrons}

\author{Antje Kiesel$^1$ \quad \quad Tobias Gauer$^2$}

\maketitle

\noindent $^1$ Institute for Mathematics, University of Rostock, Rostock, Germany\\
\noindent E-Mail: antje.kiesel@uni-rostock.de

\noindent $^2$ Department of Radiotherapy and Radio-Oncology, University Medical Center
Hamburg-Eppendorf, Hamburg, Germany\\
\noindent E-Mail: t.gauer@uke.uni-hamburg.de

\begin{abstract}
In intensity-modulated radiation therapy, optimal intensity distributions of incoming beams are decomposed into linear combinations of leaf openings of a multileaf collimator (segments). In order to avoid inefficient dose delivery, the decomposition should satisfy a number of dosimetric constraints due to suboptimal dose characteristics of small segments. However, exact decomposition with dosimetric constraints is only in limited cases possible. The present work introduces new heuristic segmentation algorithms for the following optimization problem: Find a segmentation of an approximated matrix using only allowed fields and minimize the approximation error. Finally, the decomposition algorithms were implemented into an optimization programme in order to examine the assumptions of the algorithms for a clinical example. As a result, identical dose distributions with much fewer segments and a significantly smaller number of monitor units could be achieved using dosimetric constraints. Consequently, the dose delivery is more efficient and less time consuming.
\end{abstract}



\section{Introduction}
In intensity-modulated radiation therapy (IMRT), intensity matrices with nonnegative integer entries are computed for each irradiation field. After discretization of the field into bixels, each entry of the matrix corresponds to the required intensity within this bixel. The segmentation step consists in decomposing the matrix into a linear combination of subfields (segments) shaped by a multileaf collimator (MLC). The first intuition is that a treatment plan is optimal, if the linear combination of the chosen segments equals the matrix. Such a plan consists of various segments possibly including those segments where most of the irradiation field is covered and only few bixels receive radiation.

\begin{figure}
\includegraphics[scale=0.2]{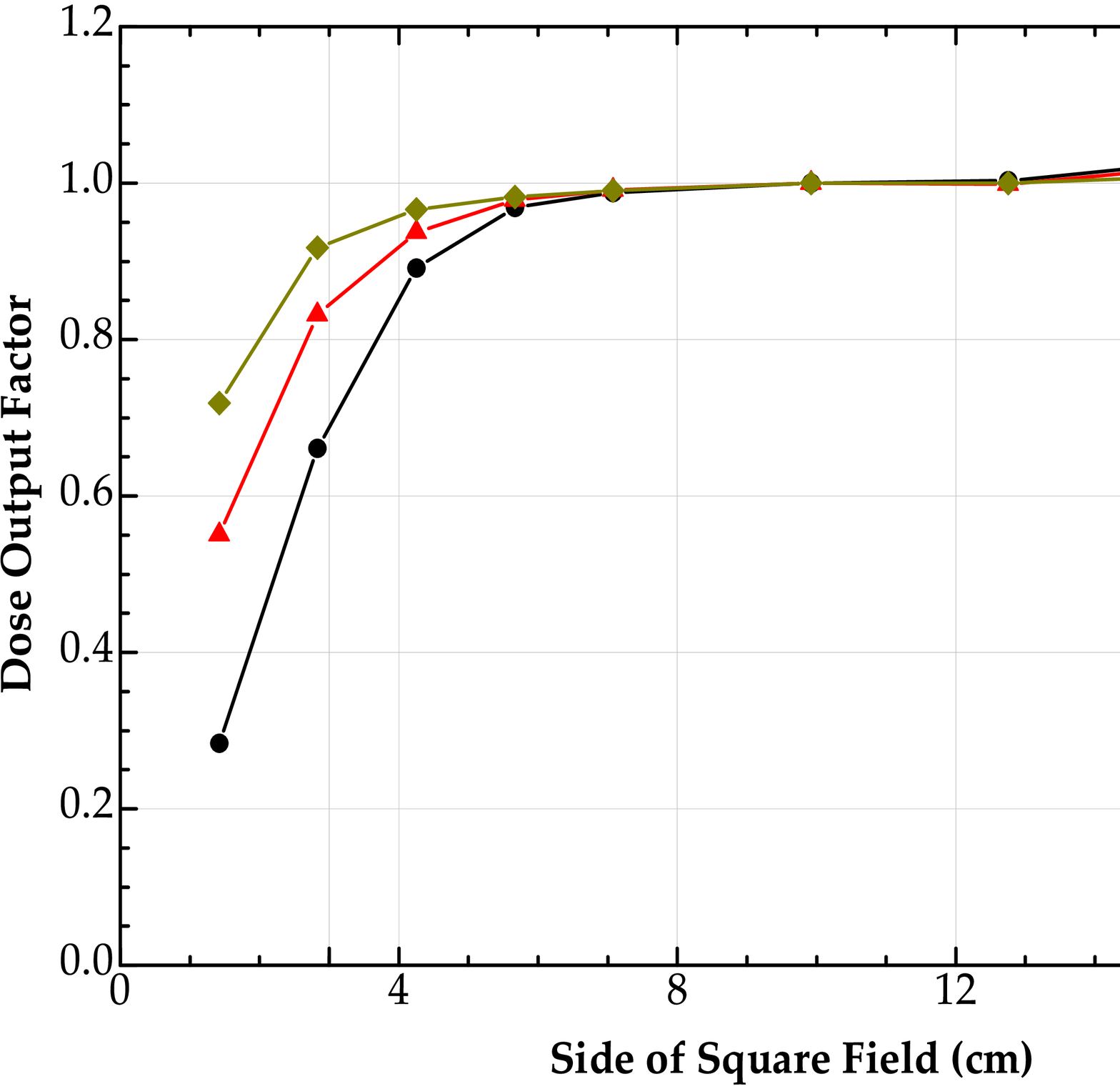}
\hspace{2em}
\includegraphics[scale=0.2]{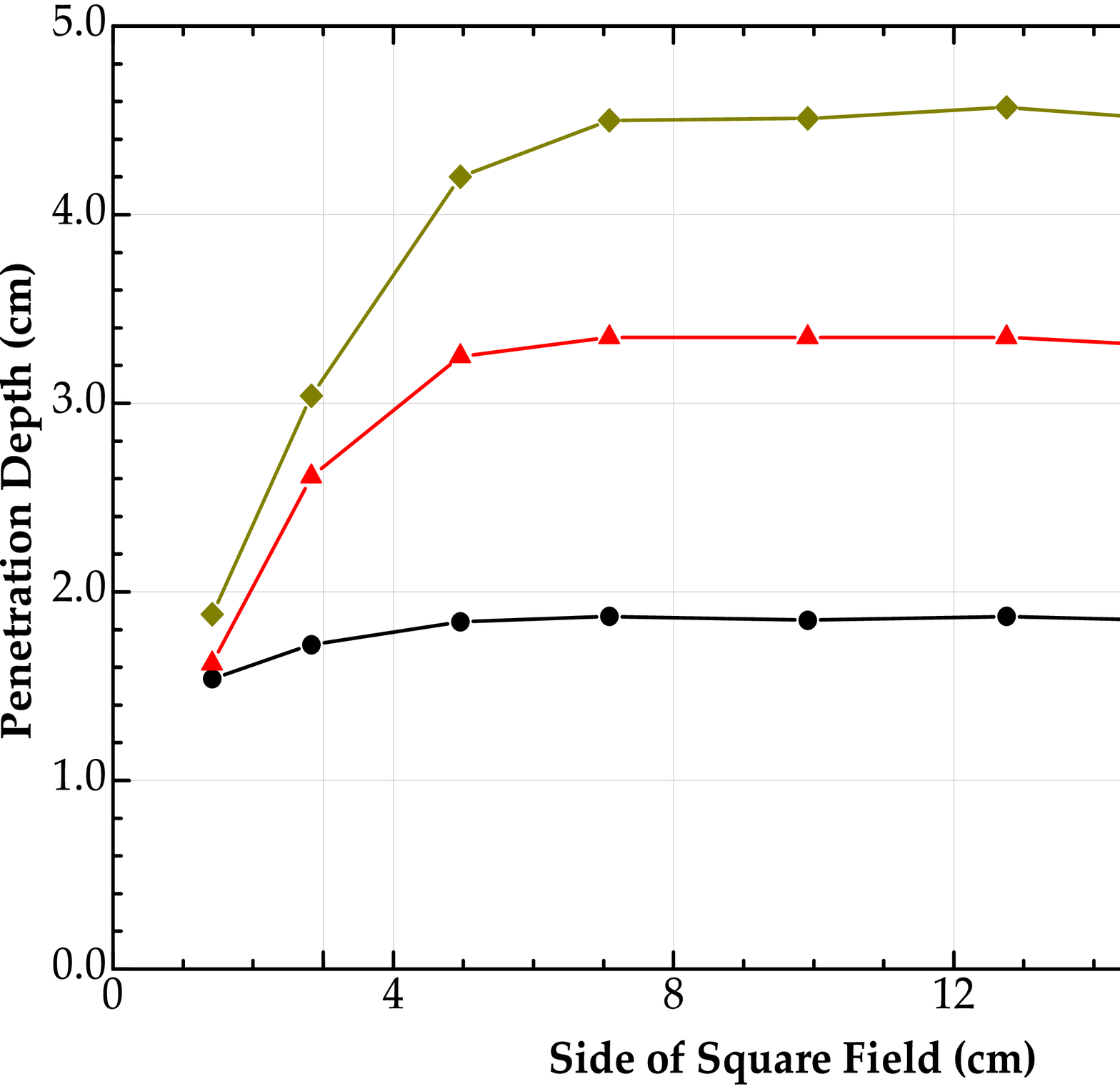}
\label{fig:properties}\caption{Electron dose output at the dose maximum normalized to the dose output of the $10$ cm $\times$ $10$ cm field and electron penetration depth of the 90 \% depth-dose as a function of square field size and electron energy (from \cite{Gau08}). The fields were shaped by an add-on MLC for electrons presented in Figure \ref{fig:mlc}. A minimum MLC field size of approximately $3$ cm $\times$ $3$ cm is necessary for decomposing intensity distributions into leaf openings to ensure an output factor of nearly $1$ and an energy-dependent penetration depth.}
\end{figure}

For dosimetric reasons, however, the model assumption is not given in practice. Irradiation of small photon or electron segments result in a much lower dose output compared to conventional conformal fields. Therefore, the linearity assumption, that irradiating one segment is equivalent to dividing it into two parts and irradiating them separately, only holds, if the two parts are still sufficiently large. In addition, the penetration depth of electrons decreases with decreasing field size and is almost independent of the beam energy for approximately $1$ cm $\times$ $1$ cm fields. However, the energy dependence of the penetration depth is necessary for our new IMRT technique with electron beams to adjust the dose to the target volume by use of various beam energies. Figure \ref{fig:properties} shows that electron fields of approximately $3$ cm $\times$ $3$ cm are necessary to keep an output factor of nearly $1$ and an energy-dependent penetration depth.

As a consequence, a treatment plan should consist of segment shapes satisfying certain constraints that ensure a minimum field size. For practical purposes it is also necessary that the field openings are connected and do not degenerate into two or more parts. Besides those dosimetric constraints, there are also technical constraints reducing the number of allowed shapes. One is the leaf overtravel constraint that accommodates the fact that the left (respectively right) leaf of the MLC cannot be shifted further than a threshold to the right (respectively left). These constraints have the consequence, that not every intensity matrix is decomposable in segments satisfying the constraints. This leads us to the task to find an approximation matrix and its decomposition into ''good'' segments, that differs from the given intensity matrix as little as possible. The aim is to generate equivalent treatment plans with good segments leading to a reduction in the segment number and monitor units, respectively.

The decomposition problem for the exact case without concerning any additional constraints is well studied. Algorithms for the minimization of the beam-on time can be found in \cite{Ahu05,Baa05,Bor94,Kal05,Kal05a,Kam03a}. Approaches for minimizing the number of used segments are given in \cite{Eng05,Kal04b,Nus06}. A variety of technical constraints are considered, see \cite{Bol04,Kal08} for the interleaf collision constraint, that prohibits an overlap of adjacent leaf pairs, and \cite{Kal08,Kal08a,Kam04a,Kam04b,Que04} for the tongue-and-groove constraint. Kamath et. al. \cite{Kam03a} also investigate the minimum separation constraint that requires a minimum leaf opening in each row and develop a criterion for a matrix being decomposable under this constraint. Engelbeen and Fiorini \cite{Engelbeen08} deal with the interleaf distance constraint where the allowed difference between two left (respectively right) leaf positions is bounded by some given threshold.

An approximation problem with the aim of reducing the total beam-on time was first formulated in \cite{EngKie08} and generalized to approximated decomposition with interleaf-collision constraint in \cite{Kal08c} and \cite{KalKie08}. The dependence between field size and output factors, penetration depth and depth-dose fall-off is outlined in \cite{Gau08}. These considerations lead to the decomposition problem using segments that satisfy some minimum field size constraints. Under these constraints, an exact decomposition of the intensity matrices is, in general, no longer possible (cf. \cite{Kam03a}) and an approximation problem has to be formulated.

Another algorithmic approach that aims at minimizing the number of segments while keeping the quality of the treatment plan is the direct aperture optimization that combines the choice of beams, apertures and weights without computing a leaf sequencing step. Shepard et al. \cite{She02} allow only a limited number of apertures for each beam, Bedford and Webb \cite{Bed06} also integrate constraints on the segment shape and size in the direct aperture optimization approach. Our algorithm is applicable if one uses intensity profile segmentation and wants to compute segmentations satisfying certain field size constraints and reducing the complexity of the plan. Matuszak et al. \cite{Mat} deal with the minimization of the monitor units by smoothing the intensity profiles.

The paper is organized as follows. Section \ref{sec:definitions} gives two definitions of what we call a segment with good dosimetric properties, one basic definition and an extended one including one further constraint. We concretely define the approximation problem and in Section \ref{sec:algorithm} propose heuristic algorithms for both definitions, each of them consisting of seven different steps. The different parts of the algorithm and their properties are analyzed in Section \ref{sec:parts}. We especially outline, that the solutions of the subproblems in step $1$ and $2$ are indeed optimal. Section \ref{sec:case} introduces the clinical case we used for testing the quality of our segmentations. Section \ref{sec:results} gives computational results for the test case and detailed numerical results for the segmentation of clinical matrices from different IMRT treatment plans.

\section{Problem formulation and definitions} \label{sec:definitions}
Throughout the paper we use the notation
\[
[k]=\{1,2,\ldots,k\}\quad\text{and}\quad[k,l]=\{k,k+1,\ldots,l\}
\]
for integers $k$ and $l$, $k \le l$. Let $A=(a_{ij})$ denote the given fluence matrix of size $m\times n$. Feasible leaf positions of the MLC are modeled as binary matrices $S=(s_{ij})$, called \emph{segments}, that satisfy the consecutive-ones-property in each row. In other words, $S$ is a segment, if there are integral intervals $[\ell_i,r_i]$ for all $i \in [m]$, representing the positions of the left and the right leaf, such that
\begin{equation}\label{eqn:segment}
s_{ij}=\begin{cases}1 \mbox{ if } l_i \le j \le r_i\\ 0 \mbox{ otherwise} \end{cases}\quad ((i,j)\in[m]\times[n]).
\end{equation}

\noindent Furthermore, for each segment $S$, we define $s_{0,j}=s_{m+1,j}=0$ for all $j \in [n]$. For the described reasons, we introduce five parameters $b_l$, $b_r$, $g_1$, $g_2$ and $f$ representing the following constraints:
\begin{enumerate}
\item[(i)] Left Leaf Overtravel Constraint: For all $i \in [m]$, we require $l_i \le b_l$. In each row the left leaf cannot be shifted more to the right than to the bixel with index $b_l$.
\item[(ii)] Right Leaf Overtravel Constraint: For all $i \in [m]$, we require $r_i \ge b_r$. In each row the right leaf cannot be shifted more to the left than to the bixel with index $b_r$.
\item[(iii)] Minimum Separation Constraint and Row Overlap: If a row $i \in [m]$ is not totally covered, we require $r_i-l_i \ge g_1-1$. Similarly, if rows $i$ and $i+1$ are not completely covered, we claim $\min(r_i,r_{i+1})-\max(l_i,l_{i+1}) \ge g_1-1$. At least $g_1$ consecutive bixels in each row receive radiation and the irradiated area of two consecutive rows overlaps in at least $g_1$ bixels.
\item[(iv)] Minimum Vertical Gap: We require a minimum vertical field size and a minimum vertical size of the covered regions, i.e. in each column consecutive ones or zeros should have a minimum number $g_2$. In detail, if $s_{i-1,j}=0$, $s_{ij}=s_{i+1,j}=\dots=s_{k-1,j}=1$ and $s_{kj}=0$ for some column $j$, we have $k-i \ge g_2$. Analogously, we require the same for consecutive zeros framed by ones.
\item[(v)] Minimum Total Field Height: At least $f$ consecutive rows of the field are not totally covered, i.e. there are at least $f$ consecutive rows with $l \le r$. This ensures, that the total size of the field is reasonably large.
\end{enumerate}
Of course, these parameters only make sense if $1 \le b_l,b_r,g_1 \le n$, $1 \le g_2,f \le m$ as well as $1 \le b_r < b_l \le n$ and $g_2 \le f$.

The case that one row of the field is totally covered and receives no radiation at all, is throughout this paper represented by the leaf positions $l=n+1$ and $r=0$. In practice, one will of course choose leaf positions of the form $l=r+1$ with $l \le b_l$ and $r \ge b_r$ that respect the leaf overtravel constraints (i) and (ii).

\vspace{1em}
{\bf Remark 1.} The Minimum Vertical Gap can be formulated in terms of the leaf position as follows:
If $l_i<l_{i-1}$ for some $i \ge 2$, then we also require $l_{i+1} \le l_i, l_{i+2} \le l_i, \dots, l_{i+g_2-1} \le l_i$. Analogously, if $r_i>r_{i-1}$ for some $i \ge 2$, we also have $r_{i+1} \ge r_i, r_{i+2} \ge r_i, \dots, r_{i+g_2-1} \ge r_i$. This ensures, that in vertical direction, we always have at least $g_2$ bixels open. Therefore, we additionally not allow that $l_i>l_{i-1}$ or $r_i<r_{i-1}$ for $2 \le i \le g_2$ and forbid also $l_i<l_{i-1}$ as well as $r_i>r_{i-1}$ for $m-g_2+1 \le i \le m$.
Similarly, we require at least $g_2$ bixels closed in vertical direction, if there are open bixels above and below in this column of the matrix. Thus, we make sure that also thin shapes in vertical direction, having negative dosimetric properties as discussed in the introduction, are forbidden.

\vspace{1em}

\begin{figure}
\centering
\input{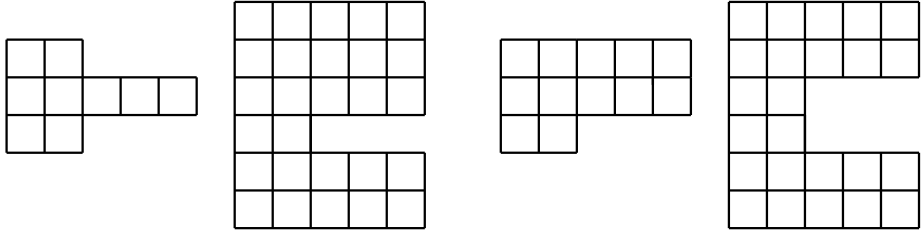} \label{fig:vertical_gap}
\caption{For $g_1=g_2=2$ the two left segments violate the minimum vertical gap constraint and only satisfy (iii), whereas the two right segments meet both conditions (iii) and (iv).}
\end{figure}

A segment is called \emph{connected} if the irradiated area that corresponds to its leaf positions does not resolve into two or more parts, i.e. if the corresponding rectilinear polygon (considered as an open set) is connected.

As the realization of the minimum vertical gap turns out to be the most difficult task, we introduce two different definitions of ``good'' segments.

\begin{defn}\label{defn:segment}
Given the parameters $b_l,b_r,g_1,f \in \mathbb{N}$ with $1 \le b_l,b_r,g_1 \le n$, $1 \le f \le m$ as well as $1 \le b_r < b_l \le n$, a \emph{segment with good dosimetric properties} $S$ is a connected segment satisfying the constraints (i), (ii),(iii) and (v).
\end{defn}

\begin{defn}
Given the parameters $b_l,b_r,g_1,g_2,f \in \mathbb{N}$ with $1 \le b_l,b_r,g_1 \le n$, $1 \le g_2,f \le m$ as well as $1 \le b_r < b_l \le n$ and $g_2 \le f$, a \emph{segment with very good dosimetric properties} is a connected segment satisfying the constraints (i)-(v).
\end{defn}

For brevity of notation we will call the segments with good dosimetric properties from now on simply segments and the segments with very good dosimetric properties advanced segments.

All in all, we have two optimization problems:
Given a matrix $A$ with positive integer entries and suitable parameters $b_l,b_r,g_1,g_2,f$, find a segmentation
\[
B=\sum\limits_{k=1}^{t} u_k S_k
\]
where the $u_k$ are positive integers and
\begin{itemize}
\item {\bf Approximated Segmentation into segments (ASS)}: the $S_k$ are segments
\item {\bf Approximated Segmentation into advanced segments (ASAS)}: the $S_k$ are advanced segments
\end{itemize}
such that
\[
\sum\limits_{i=1}^{m} \sum\limits_{j=1}^{n} |a_{ij}-b_{ij}| \rightarrow \min.
\]

The value of the objective function of the optimization problem is called \emph{total change}. The delivery time of the segmentation is $\sum_{k=1}^{t} u_k$ and the number of segments is $t$.

Obviously, for the parameter set $b_l=n$, $b_r=1$, $g_1=g_2=f=1$, segments are simply connected segments in the sense of Equation (\ref{eqn:segment}), the approximation problem has $0$ as value of the objective function and degenerates to a segmentation problem into connected segments defined by Equation (\ref{eqn:segment}).

\section{Approximated segmentation} \label{sec:algorithm}
Now we introduce two basic algorithms for {\bf ASS} and {\bf ASAS}, each consisting of seven different steps. Two of the steps are identical in both algorithms as they are computed before the segmentation step and therefore do not affect the parameter $g_2$. Five of the steps differ subject to whether the constraint $g_2$ is regarded or not (namely steps $3$-$7$). Steps $1$ and $2$ will be solved exactly, whereas steps $3$-$7$ are heuristic.

The basic structure of our algorithms is as follows:
\begin{enumerate}
\item Solve the {\bf Leaf Overtravel Constraint Problem (LOC)} on $A$: Given $b_l,b_r \in \mathbb{N}$ with $1 \le b_l < b_r \le n$, find an approximation $B$ with nonnegative integer entries that can be decomposed with respect to the Leaf Overtravel Constraint such that the total change is minimal.

    $A:=B.$
\item Solve the {\bf Minimum Separation Constraint Problem (MSC)} on $A$: Given $g_1 \in [n]$, find an approximation $B$ that can be decomposed into segments with $r_i-l_i \ge g_1-1$ for all $i \in [m]$ such that the total change is minimal.

    $A:=B.$

\item ASS: Compute an approximated segmentation $\mathcal{S}$ into connected segments satisfying (i)-(iii), but not necessarily (v).

      ASAS: Compute an approximated segmentation $\mathcal{S}$ into connected segments satisfying (i)-(iv), but not necessarily (v).

      The approximated segmentation may violate (v) and belongs to an approximation matrix $B$ that might have a large total change.
\item {\bf Combination of Fields}: Combine stepwise two disjoint fields, if this is possible with small total change. For ASAS, make sure that the new field satisfies (iv).
\item  {\bf Make-two-of-two}: For $S,S' \in \mathcal{S}$, where $S$ violates (v) and $S'$ satisfies (v), compute a substitution $S+S'=\hat{S}+\hat{S}'$, such that $\hat{S}$ and $\hat{S}'$ satisfy (v). For ASS, make sure that $\hat{S}$ and $\hat{S}'$ satisfy (i)-(iii). For ASAS, $\hat{S}$ and $\hat{S}'$ must also satisfy (iv).
\item {\bf Handle Critical Segments}: If there are still segments violating (v), try to combine them with a feasible segment such that the total change of the combination is smaller than omitting the critical segment. Take care, that (i)-(iii) (resp. (i)-(iv)) hold.
\item {\bf Total Change Improvement}: For all segments and all rows $i$, check whether an increase or decrease of $l_i$ (respectively $r_i$) reduces the total change. For ASS, change $l_i$ or $r_i$ only if (i)-(iii) still hold. For ASAS, change $l_i$ or $r_i$ only if (i)-(iv) still hold. Look at all the segments cyclically until no more changes are possible.
\end{enumerate}

The output of the algorithm is a segmentation of an approximation matrix that consists only of segments for ASS and only of advanced segments for ASAS. The LOC- and the MSC-approximation aim at producing a first approximation that can be better decomposed with the given constraints (i)-(iii) in the segmentation step than the initial matrix. The combination of fields, the make-two-of-two-step and the handling of critical segments try to provide segments satisfying (v) without producing too much total change. Finally, the total change improvement is computed in order to improve the approximation.

A very important feature of our heuristic is that further constraints can easily be taken into consideration. For example, if adjacent left and right leafs may not overlap (interleaf collision constraint) one can allow only leaf positions that respect this constraint in the optimization steps $3$-$7$.

\section{Subproblems} \label{sec:parts}
Now we describe in detail the subalgorithms and outline some of their basic properties. Throughout the steps that follow the leaf overtravel approximation, we allow only leaf positions that respect the leaf overtravel constraint. For simplicity, we will not mention this basic fact in every step.

\subsection{Leaf Overtravel Constraint}
As the leaf overtravel constraint only affects a single row of the matrix, the problem LOC can be solved for each row independently. Thus, we compute an optimal approximation of a vector $\bm{a}$. Segmentations reduce to sums of intervals $[l,r]$. Segments are simply $0$-$1$-vectors $\bm{s}$ with consecutive ones.

\begin{lem}\label{lem:loc}
A vector $\bm{a}$ has a segmentation $\bm{a}=\sum\limits_{i=1}^{k} \bm{s}_i$ with corresponding leaf positions $l_i \le b_l$ and $r_i \ge b_r$ iff $a_j \ge a_{j+1}$ for all $j \in [b_l,n-1]$ and $a_j \ge a_{j-1}$ for all $j \in [2,b_r]$.
\end{lem}
\begin{proof}
Let $a_0=a_{n+1}:=0$ and $a_+:=\max\{0,a\}$. On the one hand, the algorithm of Bortfeld (see \cite{Bor94}) provides a segmentation where the left leaf position is $j$ for exactly $(a_j-a_{j-1})_+$ segments. Analogously, the right leaf position is $j$ for $(a_j-a_{j+1})_+$ segments and no other leaf positions occur. On the other hand, it is obvious that if $a_j>a_{j-1}$ (respectively $a_j>a_{j+1}$) there will be a segment with left (respectively right) leaf position $j$ in every segmentation. This concludes the proof.
\end{proof}

Therefore, we have to find an approximation vector, that has no up-steps after index $b_l$ and no down-steps before index $b_r$. As we assume $b_r<b_l$, we can use symmetry to solve the approximation problem for the right leaf positions. Besides, the criterion from Lemma \ref{lem:loc} shows, that $b_j=a_j$ for $j \in [b_r+1,b_l-1]$ for each optimal solution of LOC. We simply need to solve the following problem for the subvector $(a_{b_l},\dots,a_n)$ and the left overtravel constraint:

\vspace{1em}
{\bf LOC-left:} Given a vector $\bm{v}=(v_1,\dots,v_k)$, find an approximation vector $\bm{w}$ with $w_j \ge w_{j+1}$ for $j \in [k-1]$ such that $\parallel \bm{v}-\bm{w} \parallel_1 = \sum\limits_{j=1}^{k} |v_j-w_j| \rightarrow \min$.

The algorithm for solving the problem LOC-left is described in Algorithm \ref{alg:loc} in the appendix. It uses a graph theoretical approach and computes a shortest path in a layered digraph, where the $j$-th layer consists of nodes representing the possible entries of the $j$-th component of the approximation vector. The problem LOC-left is similar to the Monotone Discrete Approximation Problem (MDAP) formulated in \cite{EngKie08} and the algorithm follows the same idea.

Let $min:=\min_{j \in [k]} v_j$ and $max:=\max_{j \in [k]} v_j$ and let $tc_{ij}$ be the objective value of an optimal solution of LOC-left with $w_j=i$. Let $pre_{ij}$ be the corresponding predecessor $w_{j-1}$. With respect to Algorithm \ref{alg:loc} (that uses the notation above) we yield the following

\begin{thm}\label{thm:loc}
Algorithm \ref{alg:loc} computes an optimal solution of LOC-left.
\end{thm}
\begin{proof}
The initial values $tc_{i1}$ are trivially correct. Let now $j>1$ and let $(w_1,\dots,w_j)$ be an optimal approximation of $(v_1,\dots,v_j)$ with $w_j=i$. By induction, $tc_{w_{j-1},j-1}$ is computed correctly and thus
\[
\sum\limits_{l=1}^{j} |v_l-w_l| =tc_{w_{j-1},j-1}+|v_j-i| \ge tc_{ij}.
\]
Therefore $tc_{ij}$ is a lower bound for the total change. The choice of $i_{opt}$ makes sure, that the optimal value of $w_k$ is chosen and obviously, the approximation vector from Algorithm \ref{alg:loc} realizes the lower bound for the total change of $tc_{i_{opt},k}$.
\end{proof}

\subsection{Minimum separation constraint}
Like the Overtravel Constraint, the Minimum Separation Constraint can be handled independently for each row of the matrix. Thus the task is the following:

\vspace{1em}
{\bf MSC-Row} Given a vector $\bm{v}=(v_1,\dots,v_n)$ with nonnegative integer entries, find an approximation vector $\bm{w}$ with nonnegative integer entries, such that $\bm{w}$ has a decomposition into intervals of length $\ge g_1$ and $\parallel \bm{v}-\bm{w} \parallel_1 = \sum\limits_{j=1}^{n} |v_j-w_j| \rightarrow \min$.

We know from Kamath et al. (see \cite{Kam03a}) that a vector $\bm{a}$ can be decomposed without violating the minimum separation constraint, if the optimal decomposition of their algorithm SINGLEPAIR does not violate the minimum separation constraint. For example, the vector $\bm{a}=(1,2,1)$ cannot be decomposed with $g_1=3$, as the optimal decomposition is $(1,2,1)=(1,1,0)+(0,1,1)$ and the used intervals do not have a minimum length of $3$. This motivates the approximation problem defined above.

Obviously, the problem MSC-Row can be formulated as an integer linear programming problem as follows:
\begin{eqnarray*}
\sum\limits_{j=1}^{k} \sum\limits_{j'=\min(k,j+g_1-1)}^{n} u_{j,j'} - \gamma_k & \le a_k & k \in [n]\\
-\sum\limits_{j=1}^{k} \sum\limits_{j'=\min(k,j+g_1-1)}^{n} u_{j,j'} - \gamma_k & \le -a_k & k \in [n]\\
u_{j,j'} & \ge 0 & j,j' \in [n], j' \ge j+g_1-1 \\
u_{j,j'} & \in \mathbb{Z} & j,j' \in [n], j' \ge j+g_1-1 \\
\gamma_k & \in \mathbb{Z} & k \in [n]\\
\sum\limits_{k=1}^{n} \gamma_k & \rightarrow \min
\end{eqnarray*}

We solve this integer program for each row of the matrix using SCIP \cite{Achterberg2007} with SoPlex \cite{Wunderling} as LP solver. The problem can also be solved by a combinatorial algorithm using a minimum cost flow formulation which is shown in \cite{Eng_Fio_Kie09}.

\subsection{Segmentation}
Let $B$ be the approximation matrix resulting from the MSC-step. The basic idea of the segmentation is to consider the current total change to be the sum of the absolute values of the entries of $B$ and to iteratively compute a segment $S$ whose subtraction reduces the current total change. In each step, the matrix $B$ is updated by setting $B:=B-S$. At the end of the segmentation, a positive entry in $B$ represents a bixel with underdose and a negative entry a bixel with overdose.

\subsubsection{Segmentation for ASS}
Let $\bm{b}_i$ denote the $i$-th row of $B$ for all $i \in [m]$. A segmentation consists of segments $S$ each represented by its leaf positions $l_i,r_i$ for $i \in [m]$. The main body for the segmentation step is described in Algorithm \ref{alg:seg_ASS} in the appendix. This algorithm uses the subroutine {\bf Find interval ASS} that is precisely described in Algorithm \ref{alg:find_interval_ASS}.

The idea behind this heuristic choice of the segment $S$ being subtracted from $B$ in each step is, that we compute the first interval from a sliding window segmentation (see again \cite{Bor94}), with $l$ as index of the first up-step and $r$ as index of the first down-step in the corresponding row. If these values already satisfy all requirements, we stop. Otherwise, we lengthen the interval by changing $B$, such that the overlap with the previous row increases. In order to keep the total change small, we neglect this approximation and close the row, if there is no overlap at all.

The segmentation resulting from Algorithm \ref{alg:seg_ASS} satisfies the constraints (i)-(iii) and the connectedness, but may contain segments that do not have the minimum total field height $f$.

\subsubsection{Segmentation for ASAS}
The segmentation step for ASAS differs a little bit from the ASS segmentation. For ASS, we always find a segment that has its first nonzero row exactly in that row where the current matrix $B$ has its first nonzero row. Going through the rows, we add further ones to the segment if the current sliding window interval overlaps with the previous row.

Computational tests have shown that for ASAS a different technique makes sense because the vertical criterion (iv) plays a role. In detail, whenever we decide for $l_{i+1}<l_i$ (respectively $r_{i+1}>r_i$), this immediately implies $l_{i+k}<l_i$ (respectively $r_{i+k}>r_i$) for $k=2,\dots,g_2$. Additionally, as we need at least $g_1$ consecutive ones in each row, we also know $r_{i+k} \ge \max(l_i-1,l_{i+1}+g_1-1)$ (respectively $l_{i+k} \le \min(r_i+1,r_{i+1}-g_1+1)$). Therefore, we use a matrix $S=(s_{ij})$ to store unavoidable ones, i.e. if $l_{i+1}<l_i$, we put
\begin{eqnarray}\label{eqn:1}
s_{i+k,j}=1 \mbox{ for } 2 \le k \le g_2, \ l_{i+1} \le j \le \max(l_i-1,l_{i+1}+g_1-1)
\end{eqnarray}
and if $r_{i+1}>r_i$, we define
\begin{eqnarray}\label{eqn:2}
s_{i+k,j}=1 \mbox{ for } 2 \le k \le g_2, \ \min(r_i+1,r_{i+1}-g_1+1) \le j \le r_{i+1}
\end{eqnarray}
As we require consecutive ones in each row, we also put
\begin{eqnarray}\label{eqn:3}
s_{i+k,j}=1 \mbox{ for } 2 \le k \le g_2, \ l_{i+1} \le j \le r_{i+1}
\end{eqnarray}
if $l_{i+1}<l_i$ and $r_{i+1}>r_i$ at the same time.

We also have to take care that the covered regions have a vertical minimum size. If $l_{i+1}>l_i$ (respectively $r_{i+1}<r_i$), we analogously put unavoidable zeros into our matrix $S$ using the corresponding rules to (\ref{eqn:1}) and (\ref{eqn:2}). (\ref{eqn:3}) is not necessary here, as zeros do not have to be consecutive in the rows.

\vspace{1em}
{\bf Example 1.} Let $g_1=g_2=3$ and let the $(*)$-entries of the matrix above denote the open bixels for row $1$ and $2$. Before choosing the leaf positions for row $3$, there are some unavoidable ones and zeros that have to be respected.
\[
\begin{pmatrix}
  & * & * & * & * & & &\\
* & * & * & * &   & & &\\
1 & 1 & 1 &   & 0 & & &\\
1 & 1 & 1 &   & 0 & & &\\
\end{pmatrix}
\]
\vspace{1em}

Thus, the choice of the leaf positions in one row produces unavoidable ones or zeros in other rows. Our algorithm will choose $l_i$ and $r_i$ such that the total change of this row and the corresponding unavoidable ones is minimal.
Therefore it might happen, that we do not use the first nonzero row and we also do not use the sliding window technique anymore, as the minimum vertical gap constraint (iv) prohibits so many leaf positions that we can compare the remaining ones with regard to the resulting total change. Algorithm \ref{alg:seg_ASAS} and \ref{alg:find_interval_ASAS} in the appendix show the corresponding segmentation steps.

The idea behind this algorithm is, that for each segment and in each row we look at all feasible leaf positions. For each pair $(l,r)$, we compute the value $benchmark$, which is the difference between the number of positive entries and the number of nonpositive entries in this row as well as in the corresponding unavoidable ones. The larger this value is, the better the pair $(l,r)$ suits to the segmentation. The unavoidable zeros are not taken into account because it is not necessarily bad if an entry $b_{i,j}>0$ is closed, as this entry can be part of the following segments. We close a row, if the corresponding optimal value of benchmark is zero.

Finally, our procedure leads to a segmentation satisfying (i)-(iv) and the connectedness, but not necessarily satisfies the minimum total field height $f$.

\vspace{1em}
{\bf Remark 4.} One might argue that it is possible for ASS to compute exactly the segment, that reduces the total change in this step as much as possible. For example, one can consider a layered digraph with $m$ layers of nodes. In layer $i\in [m]$, we have nodes $(i,l,r)$ representing feasible leaf positions and we draw an edge between $(i,l,r)$ and $(i+1,l',r')$, if the combination of these two leaf positions satisfies all constraints. Furthermore, we draw edges from a source to all nodes in the first layer and from all nodes in the last layer to a sink. The edge weights are just the total change reductions caused by the leaf positions of the end node of the edge and zero for all edges whose end node is the sink. For a detailed description of the graph see \cite{Bol04}. The optimal segment can then be found by shortest path computation in the digraph. But indeed, such a choice is not a good idea because reducing the total change as much as possible leads to badly decomposable residual matrices. For example, for $g_1=2$, $\bm{b}=(1,2,1)$ would be reduced by $(1,1,1)$ and the residuum $(0,1,0)$ is badly decomposable. Our used sliding-window-technique is better and leads to $(1,2,1)=(1,1,0)+(0,1,1)$.

For ASAS the constraints are too complex anyway to compute the optimal segment in one step, that reduces the total change by a maximum value.

\subsection{Combination of fields}
Given two segments $S=((l_1,r_1),\dots,(l_m,r_m))$ and $S'=((l'_1,r'_1),\dots,(l'_m,r'_m))$, let us consider the open regions $i_1,\dots,i_2$ and $i'_1,\dots,i'_2$ of $S$ and $S'$, precisely
\begin{eqnarray}
\{i_1,\dots,i_2\}=\{i \in [m] \ : \ l_i<n+1\} \label{open1}\\
\{i'_1,\dots,i'_2\}=\{i \in [m] \ : \ l'_i<n+1\} \label{open2}
\end{eqnarray}
If $i'_1=i_2+1$ and $\min(r_{i_2},r'_{i'_1})-\max(l_{i_2},l'_{i'_1}) \ge g_1-1$, we merge $S$ and $S'$ and get one new segment $S''$ with
\[
s''_{ij}=
\begin{cases}
1, & \mbox{ if } s_{ij}=1 \mbox{ or } s'_{ij}=1,\\
0, & \mbox{ otherwise.}
\end{cases}
\]

We iterate this step, until no two segments can be merged by this procedure. Obviously, this step does not affect the total change.

Afterwards, we compute a second combination step and merge segments if $i'_1=i_2+2$ and $\min(r_{i_2},r'_{i'_1})-\max(l_{i_2},l'_{i'_1}) \ge g_1-1$. This means, there is only one closed row between the two segments. We compute leaf positions $l \in [\max(l_{i_2},l'_{i'_1}),\min(r_{i_2},r'_{i'_1})-g_1+1]$ and $r \in [l+g_1-1,\min(r_{i_2},r'_{i'_1})]$ such that putting ones to the interval $[l,r]$ in row $i_2+1$ produces the smallest total change with respect to the current approximation matrix.

Again, we drop $S$ and $S'$ out of our segmentation and this time add $S''$ with
\[
s''_{ij}=
\begin{cases}
1, & \mbox{ if } s_{ij}=1 \mbox{ or } s'_{ij}=1 \mbox{ or } \\
   & \quad (i=i_2+1, l \le j \le r),\\
0, & \mbox{ otherwise.}
\end{cases}
\]

Again, we iterate this procedure, until no more such merges are possible. Obviously, this second combination step affects the total change, as we increase the approximation matrix by adding ones to the segments. After both of our combination steps, all segments still satisfy (i)-(iii), while even more segments satisfy (v) now. Another positive consequence is a reduction of the total number of (not necessarily pairwise different) segments, called the Delivery Time. The combination step is demonstrated in Figure 3.

For ASAS, we only combine two segments according to one of the two steps described above, if the criterion with the minimum vertical zeros (iv) is not violated after the combination.

\begin{figure}
\input{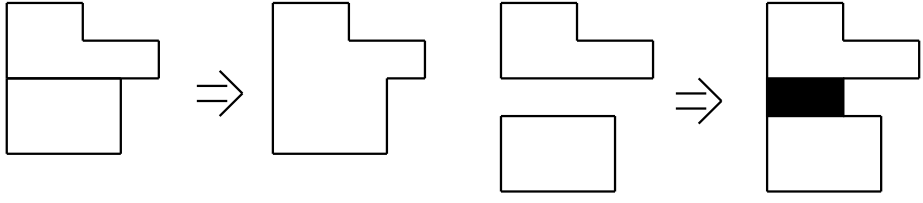}\label{fig:combination}
\caption{The first and the second combination step for ASS with $g_1=2$. The black rectangle produces an increase of the total change.}
\end{figure}

\subsection{Make-two-of-two}
We now define a substitution step $S+S'=\hat{S}+\hat{S}'$. For ASAS, the same step is computed if $\hat{S}$ and $\hat{S}'$ still satisfy (iv).

Let us again consider a segment $S=((l_1,r_1),\dots,(l_m,r_m))$ and its open region $i_1,\dots,i_2$ defined by (\ref{open1}). If $S$ violates (v), we call $S$ a critical segment. Now we check whether we find a segment $S'=((l'_1,r'_1),\dots,(l'_m,r'_m))$ with open region $i'_1,\dots,i'_2$  defined by (\ref{open2}) such that
\begin{eqnarray*}
i_2-i'_1 \ge f-1 \mbox{ and } i'_2-i_1 \ge f-1.
\end{eqnarray*}
Note that in this case, we have $i'_1<i_1$ and $i'_2>i_2$ due to $i_2-i_1<f-1$. Thus, the set of open rows of $S$ is a subset of the set of open rows of $S'$. If
\begin{eqnarray}
\min(r_{i_1},r'_{i_1-1})-\max(l_{i_1}-l'_{i_1-1}) \ge g_1-1, \label{two1}
\end{eqnarray}
we substitute $S$ and $S'$ by segments $\hat{S}=((\hat{l}_1,\hat{r}_1),\dots,(\hat{l}_m,\hat{r}_m))$ and \newline $\hat{S}'=((\hat{l}'_1,\hat{r}'_1),\dots,(\hat{l}'_m,\hat{r}'_m))$ defined as follows
\begin{eqnarray*}
(\hat{l}_i,\hat{r}_i)=
\begin{cases}
(l'_i,r'_i) & \mbox{  if } i < i_1,\\
(l_i,r_i) & \mbox{  if } i \ge i_1,
\end{cases}
\quad
(\hat{l}'_i,\hat{r}'_i)=
\begin{cases}
(l_i,r_i)=(n+1,0) & \mbox{  if } i < i_1,\\
(l'_i,r'_i) & \mbox{  if } i \ge i_1.
\end{cases}
\end{eqnarray*}

The result is that we add the upper part of segment $S'$ to segment $S$ in order to enlarge $S$, while $S'$ remains sufficiently large. If (\ref{two1}) is not satisfied, we can have a second try and check whether
\begin{eqnarray}
\min(r_{i_2},r'_{i_2+1})-\max(l_{i_2}-l'_{i_2+1}) \ge g_1-1 \label{two2}.
\end{eqnarray}
If this condition is true, we can analogously add the lower part of $S'$ to $S$ and close all rows $\ge i_2+1$ of $S'$. The Make-two-of-two-procedure is illustrated in Figure 4.

\begin{figure}
\input{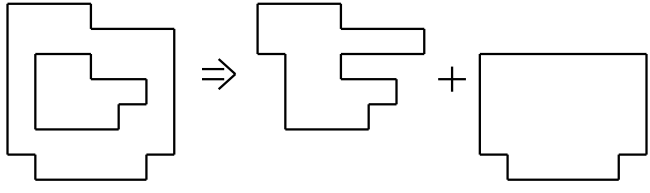}\label{fig:make_two_of_two}
\caption{The Make-two-of-two-step for ASS with $g_1=2$ and $f=4$.}
\end{figure}

The make-two-of-two-step is computed for all critical segments. If we find a partner, we compute the substitution immediately. If no partner is found throughout the segmentation, the segment $S$ is dropped and stored in a new list of critical segments violating (v). Whereas the substitution with a partner does not affect the total change, the elimination of segments from the segmentation that find no partner leads to an increase of the total change.

\subsection{Handle Critical Segments}
Let $S=((l_1,r_1),\dots,(l_m,r_m))$ be a critical segment stored in the make-two-of-two-step and let $i_1,\dots,i_2$ be its open region defined by (\ref{open1}). We try to combine $S$ with a partner $S'=((l'_1,r'_1),\dots,(l'_m,r'_m))$ from the segmentation with open region $i'_1,\dots,i'_2$ defined by (\ref{open2}) if this combination step causes lower total change then simply omit $S$.

More precisely, if $i'_2 \ge i_2$ and $i'_1 \le i_1$, we define a new segment $S''=((l''_1,r''_1),\dots,(l''_m,r''_m))$ with
\[
(l''_i,r''_i)=
\begin{cases}
(\min(l_i,l'_i),\max(r_i,r'_i)), \mbox{ if } i_1 \le i \le i_2,\\
(l'_i,r'_i), \mbox{ otherwise.}
\end{cases}
\]

The resulting segment $S''$ emanates from $S'$ by attaching certain (or all) parts of $S$ and possibly also adding connecting ones. We now drop $S'$ from our segmentation and $S$ from the list of critical segments and add $S''$ to the segmentation, if the total change that is caused by this decision is smaller than simply omitting $S$. If no partner for $S$ is found, we simply delete $S$ from the list of critical segments and accept the corresponding total change.

For ASAS, we execute the step if (iv) is not violated.

\begin{figure}
\input{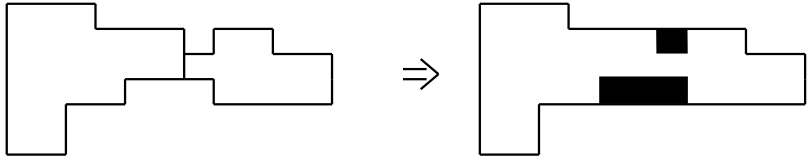}\label{fig:critical_segments}
\caption{The Handle-critical-segments-step for ASS with $g_1=2$ and $f=4$. The black rectangle shows the combining bixels increasing the total change.}
\end{figure}

\subsection{Total Change Improvement}
As the previous segmentation steps are indeed not optimal, it may happen, that some leaf positions can be increased (respectively decreased) without violating the constraints and the total change becomes smaller.

Let $B$ be the approximation matrix that corresponds to the current segmentation. The improvement step is computed using Algorithm \ref{alg:improvement} which is described in the appendix and adjusts the segmentation such that $B$ gets closer to $A$ with regard to the total change.

\section{Clinical case}\label{sec:case}

A clinical case was set-up to examine the efficiency of our proposed segmentation algorithms. For a patient with cancer of the right breast, electron irradiation plans using various segmentation settings and different optimization constraints were created with a self-designed IMRT optimization programme based on our previous studies \cite{Eng05a}. The planning target volume was the right breast, which should receive a total dose of 50.4 Gy (1.8 Gy per fraction). In addition, the target volume should be covered by the $95\%$ isodose line ($95\%$ of the prescribed dose). The ipsilateral lung was considered to be organ at risk.

\begin{figure}[t]
\center{\includegraphics*[height=4cm]{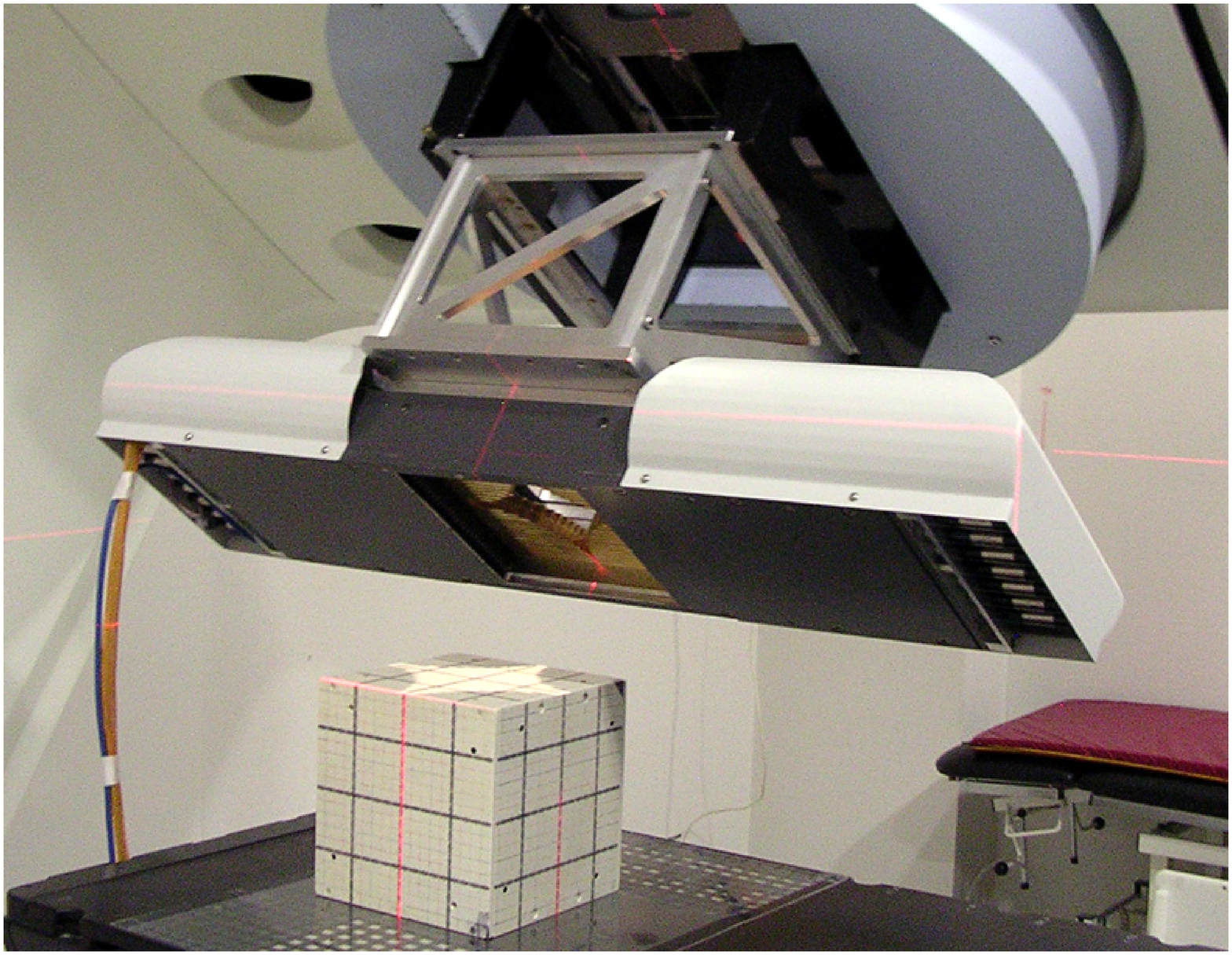}
\hskip0.3cm \includegraphics*[height=4cm]{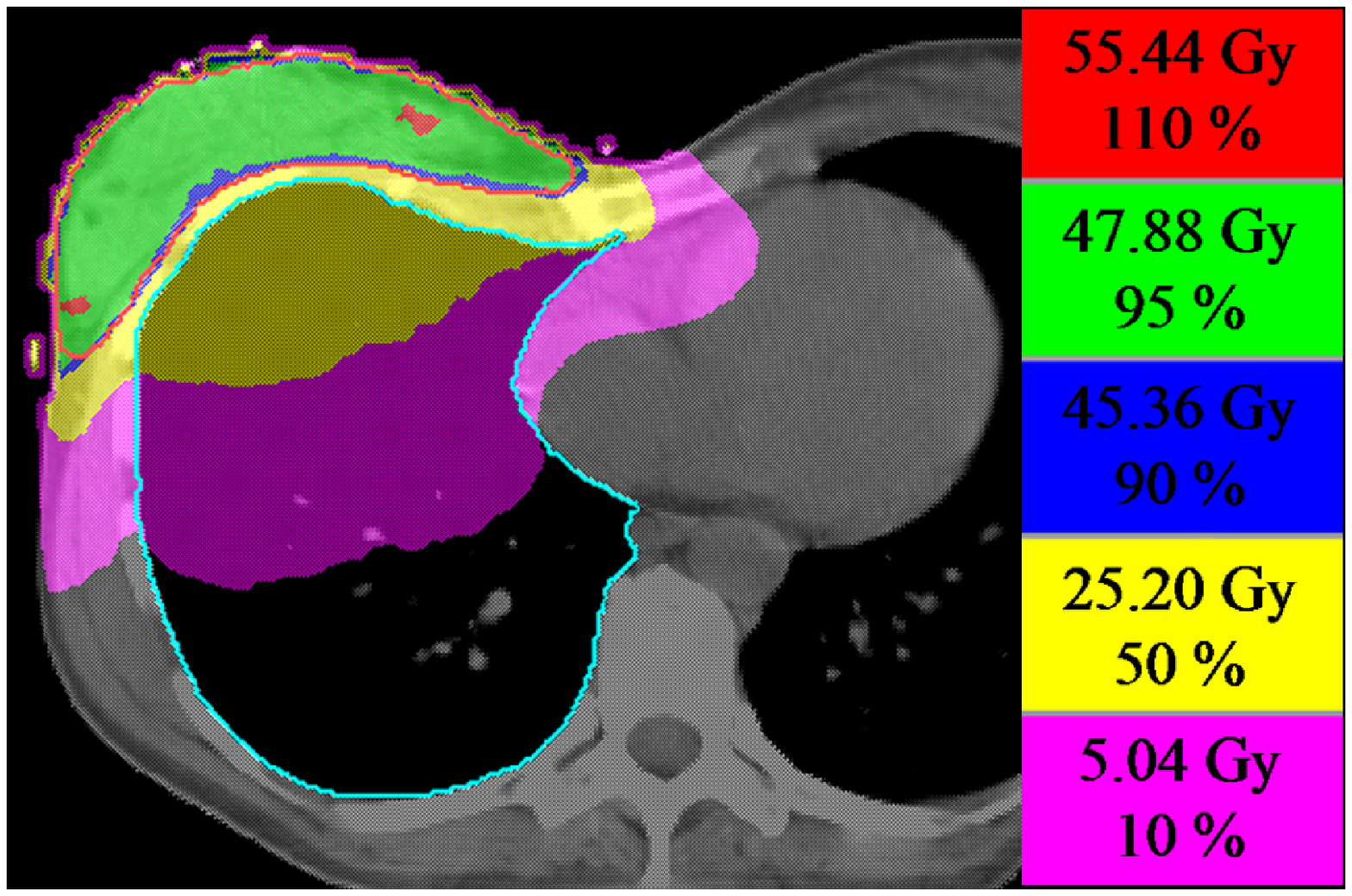}}
\caption{Left: Add-on MLC for electrons mounted on a conventional Siemens accelerator. Right: Dose distribution of an electron IMRT plan consisting of 26 MLC fields achieved through segmentation setting $442$. The corresponding dose volume histogram is shown in Figure \ref{fig:dvh} (left). The setting $442$ is given by a minimum total field height of $4$, a minimum separation constraint and row overlap of $4$ and a minimum vertical gap of $2$.}
\label{fig:mlc}\label{fig:dose}
\end{figure}

The optimization programme provides simultaneous optimization of beam orientation, energy and intensity for dose delivery with an add-on MLC for electrons (Euromechanics, Schwarzenbruck, Germany) presented in Figure \ref{fig:mlc} and \cite{Gau06, Gau08}. Electron dose calculation was performed by Monte Carlo simulations with the treatment planning system \textit{Pinnacle} from Philips (Version 8.1s). Final dose calculation of the treatment plans was conducted using a dose grid size of $3$ mm and a dose calculation uncertainty of $2\%$. The following optimization steps are necessary to generate an electron IMRT plan:

\begin{enumerate}
\item Simultaneous optimization of beam orientation, energy and intensity: A set of radiation incidence angles (typically 10--15) is determined given by table and gantry angles \cite{Eng05a,Lim07}. For each configuration, the algorithm calculates the optimal fluence distribution, given by a nonnegative integer matrix.
\item The intensity matrices are approximately decomposed into a superposition of allowed segments such that the deviation between desired and actual fluence is minimal. The result is a set of segments, each of them is given by the corresponding MLC leaf positions and its dose weight.
\item The segments from step 2 are treated as candidates for the treatment plan. In a third optimization step, the dose of the candidates are calculated for all beam energies and then optimized for a given weight proportion between best target coverage and minimum dose to critical organs in order to find the final set of segments with optimal beam energies and their corresponding monitor units.

\end{enumerate}

In this paper, we have focused on step 2 and introduced optimization algorithms for an approximate decomposition of intensity matrices.

Until now, the segmentation step consisted in exactly decomposing the intensity matrices using all deliverable segments. In our approach, we admit a decrease in the decomposition accuracy in order to obtain segments which satisfy the dosimetric and technical constraints. Step 3 justifies the approximation approach in Step 2, as a larger approximation error does not necessarily result in a suboptimal treatment plan. Indeed, larger segments produce homogeneous dose distributions and thus, the same final fluence can be generated using fewer larger segments. The acceptability of a treatment plan is decided after step 3 by means of dose volume histograms (see Section \ref{sec:results}) and a plan is only presumed if the required dose constraints are not exceeded. Therefore, the danger of cumulative deviation in the approximation step does not really exist, as the computed segments are just candidates for the treatment plan that pass through a further optimization step.

\section{Results}\label{sec:results}
At first, we compare electron IMRT plans created with different segmentation settings for the clinical case prescribed in Section \ref{sec:case}. The comparison was conducted using two different optimization settings: one setting to achieve with a better dose coverage of the breast and the other one to reach a better sparing of the lung. Note, that the segmentation settings refer to parameters in the segmentation step, which is discussed here, whereas the optimization settings play a role in the final optimization step that is not part of this paper. Finally, we give a detailed evaluation for the results of the decomposition step.

A treatment plan with a segmentation setting $xyz$ uses the decomposition algorithm with a minimum total field height of $f=x$, a minimum separation constraint and row overlap of $g_1=y$ and a minimum vertical gap of $g_2=z$. The decomposed matrices vary in their vertical size $m$ and their horizontal size $n$, as they describe only parts of the beam head where the target volume is located. Thus, in practice, the overtravel parameters $b_l$ and $b_r$ will depend on the positioning of the matrix and are put individually for each matrix. Our electron MLC is capable of shifting the leaves edges to $3/4$ of the radiation field.

\begin{figure}[t]
\center{\includegraphics*[bb=0 0 850 600,scale=0.2]{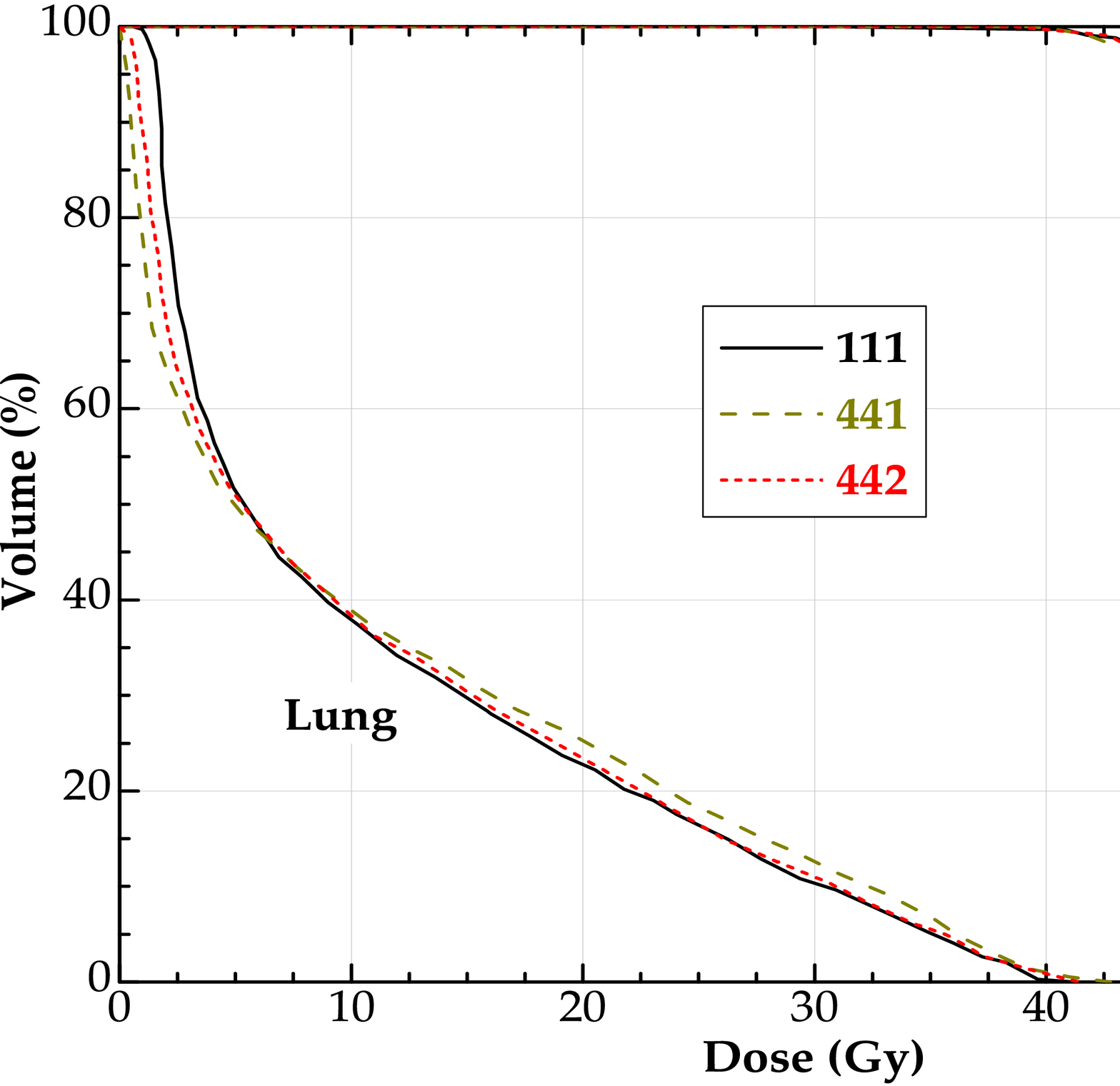}
\includegraphics*[bb=0 0 850 600,scale=0.2]{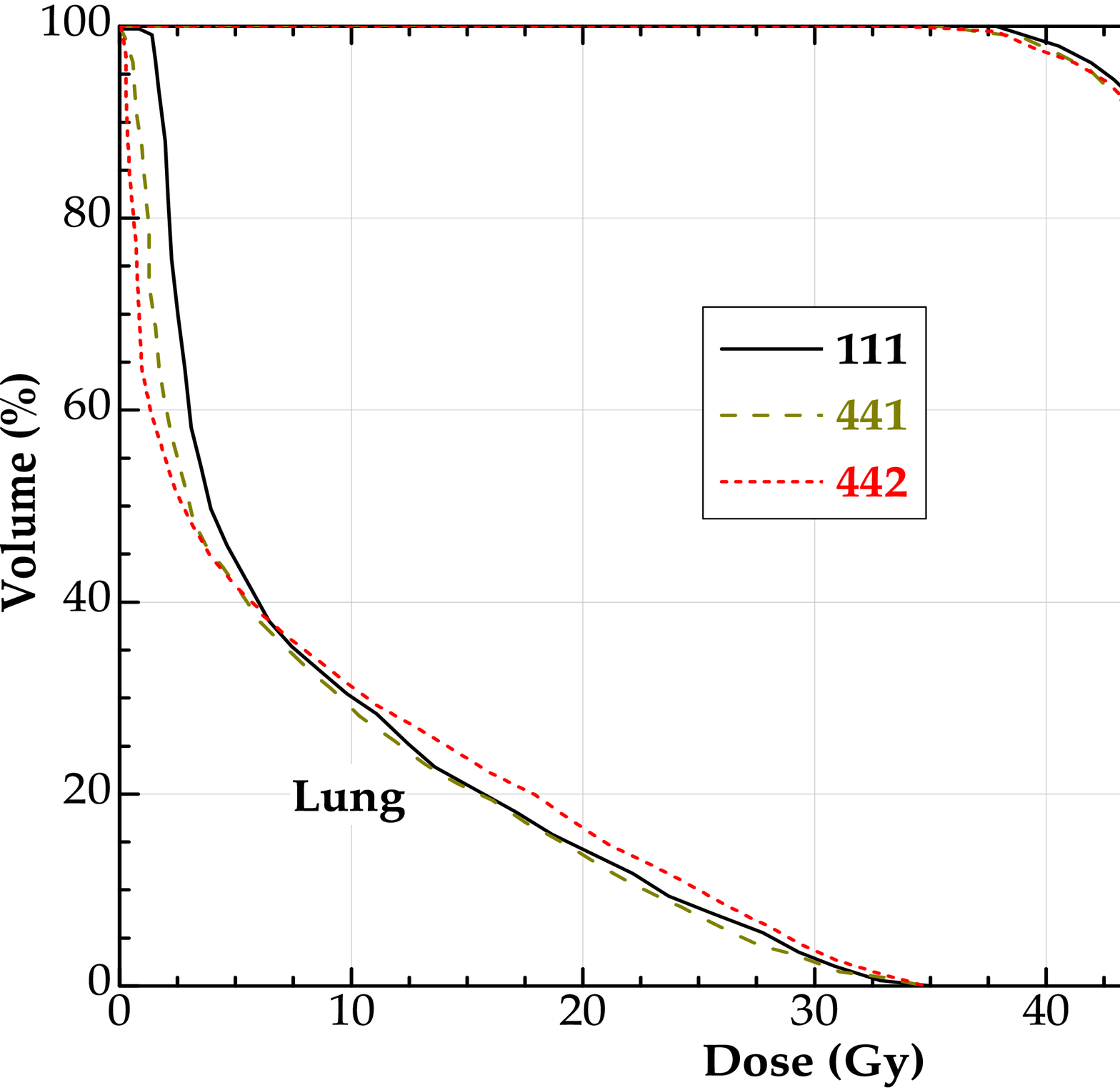}}
\caption{Dose volume histograms for settings $111$, $441$ and $442$ to demonstrate that identical results could be achieved when using greater minimum segment sizes (cf. $111$ and $441$) and segment shapes without single vertical leaf openings and leaf closings (cf. $441$ and $442$). The setting $xyz$ is given by a minimum total field height of $x$, a minimum separation constraint and row overlap of $y$ and a minimum vertical gap of $z$. Left: Dose volume histogram for optimization setting to reach a better dose coverage of the breast volume. The resulting dose distribution for setting $442$ is presented in Figure \ref{fig:dose}. Right: Dose volume histogram for optimization setting to provide a better sparing of the lung.}\label{fig:dvh}
\end{figure}

The plan quality was evaluated by means of dose volume histograms that indicate the amount of dose delivered to a certain volume of the patient (here: the right breast and the right lung). Thus, dose homogeneity in the target volume and dose exposure to the organs at risk can be examined. In Figure \ref{fig:dvh}, the dose volume histograms for both optimization settings demonstrate that almost identical dose distributions can be achieved using smaller or larger minimum MLC openings (cf. setting $111$ and $441$). In fact, the treatment plan could be slightly improved by use of a minimum vertical gap parameter of $2$ which avoids single leaf openings and closings.

Table \ref{table1} and \ref{table2} illustrate the main benefit of our approach, as identical results can be achieved with approximately two thirds fewer segments and a significantly smaller number of monitor units by use of dosimetric constraints. As a result, the dose delivery is more efficient and less time consuming. The minimum number of segments is reached for setting $442$ and computational tests have shown, that this configuration produces the optimal results. As the leaf width is $0.7$ cm, fields with a horizontal and vertical height of $4$ bixels have a size of approximately $3$ cm $\times$ $3$ cm and this confirms our dosimetric constraint of $3$ cm $\times$ $3$ cm minimum segment size (cf. Figure \ref{fig:properties}). It can be also demonstrated that minimum segment sizes greater than setting $442$ do not necessarily result in fewer segments (cf. Table \ref{table1}), although the number of segments in Table \ref{table2} is slightly lower for setting $552$. For both optimization settings, the dose volume histograms were considerably better when using minimum segment sizes smaller than setting $552$.

It is important to underline that the number of segments and the number of monitor units in Table \ref{table1} and \ref{table2} belong to the final IMRT plan and result from the third optimization step and not from the decomposition step of our algorithms. In fact, the monitor units have another scale here and are not directly comparable with the delivery time from the segmentation. In contrast, the total change information stems from the decomposition step. Note, that the total change of the segmentation itself is not a significant quantity, because if the matrix entries are large, a larger total change is acceptable. Therefore, we compute the total sum of entries for each intensity matrix and then calculate the relative total change which is the ratio between total change and total sum of entries. The smaller the relative total change, the better is the decomposition.

\begin{table}[t] \footnotesize
\caption{Segmentation results of IMRT plans using different decomposition settings for optimization setting to achieve the best target coverage. Setting $xyz$ means $f=x$ (minimum total field height), $g_1=y$ (minimum separation constraint and row overlap) and $g_2=z$ (minimum vertical gap).}\label{table1}
\vskip0.2cm
\begin{tabular*}{\columnwidth}{@{}l*{15}{@{\extracolsep{0pt plus12pt}}l}}
\hline\hline
Setting		&Mean Relative Total Change	&Number of Segments	&Number of Monitor Units\\
\hline
111		&	0.04				&90		&85586		\\
221		&	0.16				&79		&65677		\\
222		&	0.34				&54		&63461		\\
331		&	0.22				&49		&38598		\\
332		&	0.36				&32		&21265		\\
333		&	0.40				&40		&36789		\\
441		&	0.30 				&55		&34262		\\
442		&	0.41				&26		&12860		\\
443		&	0.45				&28		&13119		\\
444		&	0.49				&27		&16899		\\
551		&	0.35				&40		&19162		\\
552		&	0.46				&30		&10337		\\
\hline\hline		
\end{tabular*}
\end{table}

\begin{table}[t] \footnotesize
\caption{Segmentation results of IMRT plans using different decomposition settings for optimization setting to reach less dose to the lung. Setting $xyz$ means $f=x$ (minimum total field height), $g_1=y$ (minimum separation constraint and row overlap) and $g_2=z$ (minimum vertical gap).}\label{table2}
\vskip0.2cm
\begin{tabular*}{\columnwidth}{@{}l*{15}{@{\extracolsep{0pt plus12pt}}l}}
\hline\hline
Setting		&Mean Relative Total Change	&Number of Segments	&Number of Monitor Units\\
\hline
111		&	0.04				&78		&	59399	\\
221		&	0.16				&71		&	46627	\\
222		&	0.34				&51		&	42411	\\
331		&	0.22				&52		&	25896	\\
332		&	0.36				&29		&	16700	\\
333		&	0.40				&36		&	24957	\\
441		&	0.30				&45		&	23509	\\
442		&	0.41				&28		&	10200	\\
443		&	0.45				&28		&	9995	\\
444		&	0.49				&27		&	11318	\\
551		&	0.35				&37		&	13787	\\
552		&	0.46				&25		&	8832	\\
\hline\hline		
\end{tabular*}
\end{table}

For the detailed evaluation of our algorithms, we use a set of $264$ clinical intensity matrices that originate from electron treatment plans for different patients and beam angles. The matrices are produced during the optimization step 1 of the treatment planning that was introduced in Section \ref{sec:case} and uses the algorithm from \cite{Eng05a}. Exemplarily, we compute segmentations for the settings $f=3$, $g_1=3$ and $g_2=1$ as well as $f=4$ and $g_1=g_2=2$. The values of the overtravel parameters are also produced in the pre-segmentation step. The results are shown in Table \ref{table3} and \ref{table4} and demonstrate how much total change is caused respectively avoided by the steps of the algorithms. The overtravel-approximation and MSC-approximation lead to a certain total change of the matrix that is put into the segmentation step. As an exact decomposition in the segmentation step is impossible, the total change increases here again. Both the combination step and the make-two-of-two step try to eliminate segments not satisfying the parameter $f$ and again cause some total change. Finally, the last two steps of the algorithm improve the performance and reduce the approximation error as much as possible. One can see that the combination step and the make-two-of-two-step are performed more often for ASS, as for ASAS the vertical gap ensures that the fields already have a reasonable size after the segmentation step. Of course, the larger the parameters and thus the minimum field size, the larger becomes the total change.

The first column in Table \ref{table3} and \ref{table4} gives the average results over the $264$ matrices, while the second (respectively third) column represent the single results for the matrix with the smallest (respectively largest) relative total change. Homogeneous matrices with large nonzero areas can be decomposed quite well, while matrices with only few nonzero entries that do not span connected areas lead to unacceptable results. As a treatment plan is a superposition of several intensity profiles from different beam angles, the approximation errors balance each other and lead to applicable treatment plans as described above. Furthermore, a certain part of the total change is unavoidable if one requires the constraints (i)-(v), e.g. the total change after MSC-approximation is a good lower bound for the achievable total change. All in all, taking the vertical gap $g_2$ into account increases the total change while reducing the number of used segments and the monitor units.

\begin{table}[t] \footnotesize
\caption[Numerical results for ASS]{Numerical results for ASS with parameters $g_1=3$ and $f=3$.}\label{table3}
\vskip0.2cm
\begin{tabular*}{\columnwidth}{@{}l*{15}{@{\extracolsep{0pt plus12pt}}l}}
\hline\hline
					&Average 		&Rel. TC min.	&Rel. TC max.\\
\hline
m                     &17.55 	 	&20			&11 \\
n                     &21.44  		&19 			&13 \\
Total sum of entries  &886.4  		&1329  		&152\\
Total change          &82.24  		&12  			&64\\
Delivery time         &15.39		&13			&3\\
Number of segments    &15 		&13			&3\\
\hline
TC after Overtravel-Approximation  	&26.34		&6			&75\\
TC after MSC-Approximation          &57.09		&11			&88\\
TC after Segmentation               &74.33		&13			&91\\
TC change after combination         &80.96   	&28			&91\\
TC after make-two-of-two           	&93.77		&28			&101\\
TC after handle critical segments 	&88.72		&28			&94\\
TC after improvement              	&82.24		&12			&94\\
\hline
Combinations                        &9.19 		&5			&0\\
Successful make-two-of-two          &2.78	    &0	    	&0\\
\hline
Relative total change               &0.16      	&0.009 		&0.62\\
\hline\hline		
\end{tabular*}
\end{table}

\begin{table}[t] \footnotesize
\caption[Numerical results for ASAS]{Numerical results for ASAS with parameters $g_1=g_2=2$ and $f=4$.}\label{table4}
\vskip0.2cm
\begin{tabular*}{\columnwidth}{@{}l*{15}{@{\extracolsep{0pt plus12pt}}l}}
\hline\hline
					&Average 		&Rel. TC min.	&Rel. TC max.\\
\hline
m                    &17.55  		&23  			&22  \\
n                    &21.44  		&26  			&28 \\
Total sum of entries &886.4 		&997  		&384\\
Total change         &157.7  		&46  			&269 \\
Delivery time        &7.69  		&4			&4 \\
Number of segments   &6.82  		&3			&4\\
\hline
TC after Overtravel-Approximation   	&26.34  	&7			&89	\\
TC after MSC-Approximation          	&44.39  	&7			&134	\\
TC after Segmentation               	&144.8 		&46  		&267	\\
TC change after combination         	&144.8 		&46			&267	\\
TC after make-two-of-two            	&159.2		&46 		&270 	\\
TC after handle critical segments   	&158.8 		&46			&270	\\
TC after improvement                	&157.7 		&46			&269	\\
\hline
Combinations                        	&0.24			&0			&0\\
Successful make-two-of-two          	&0.57			&0			&0\\
\hline
Relative total change               	&0.27 			&0.05  		&0.70\\
\hline\hline		
\end{tabular*}
\end{table}

\section{Conclusion}
In the present study, dosimetric and technical constraints have been taken into consideration in intensity-modulated radiation therapy (IMRT). A set of $5$ parameters has been introduced, two of them for modelling the leaf overtravel constraint, the other three to ensure a minimum field size and to avoid thin field shapes. We proposed algorithms for approximate segmentation of intensity matrices using segments that satisfy the constraints. We basically distinguish between two approximation problems depending on whether the vertical gap parameter is considered or not. The objective function of the optimization is the deviation between the desired and the approximated intensity profile that has to be minimized. The segmentation step is part of an IMRT optimization process which was examined by comparisons of dose volume histograms of treatment plans with small and large segments as well as with and without thin segment shapes. The histograms show that the use of larger segments results in equal IMRT plans with fewer segments and monitor units respectively. Although the approximation error of the segmentations rises with increasing minimum field size, equivalent or even better dose distributions could be achieved. Concluding, this first approach to approximated segmentation in IMRT planning shows the potential of these ideas and there is a need for further research in related approximation problems.

\bibliographystyle{plain}
\bibliography{IMRT}

\begin{thebibliography}{10}

\bibitem{Achterberg2007}
T.~Achterberg.
\newblock {\em Constraint {I}nteger {P}rogramming}.
\newblock PhD thesis, Technische Universit{\"a}t Berlin, 2007.
\newblock \url{http://opus.kobv.de/tuberlin/volltexte/2007/1611/}.

\bibitem{Ahu05}
R.K. Ahuja and H.W. Hamacher.
\newblock A network flow algorithm to minimize beam-on time for unconstrained
  multileaf collimator problems in cancer radiation therapy.
\newblock {\em Networks}, 45(1):36--41, 2005.

\bibitem{Baa05}
D.~Baatar, H.W. Hamacher, M.~Ehrgott, and G.J. Woeginger.
\newblock Decomposition of integer matrices and multileaf collimator
  sequencing.
\newblock {\em Discrete Appl. Math.}, 152(1-3):6--34, 2005.

\bibitem{Bed06}
J.L. Bedford and S.~Webb.
\newblock Constrained segment shapes in direct-aperture optimization for
  step-and-shoot {IMRT}.
\newblock {\em Med. Phys.}, 33(4):944--958, 2006.

\bibitem{Bol04}
N.~Boland, H.~W. Hamacher, and F.~Lenzen.
\newblock Minimizing beam-on time in cancer radiation treatment using multileaf
  collimators.
\newblock {\em Networks}, 43(4):226--240, 2004.

\bibitem{Bor94}
T.R. Bortfeld, D.L. Kahler, T.J. Waldron, and A.L. Boyer.
\newblock X--ray field compensation with multileaf collimators.
\newblock {\em Int. J. Radiat. Oncol. Biol. Phys.}, 28:723--730, 1994.

\bibitem{Eng05}
K.~Engel.
\newblock A new algorithm for optimal multileaf collimator field segmentation.
\newblock {\em Discrete Appl. Math.}, 152(1-3):35--51, 2005.

\bibitem{EngKie08}
K.~Engel and A.~Kiesel.
\newblock Approximated matrix decomposition for {IMRT} planning with multileaf
  collimators.
\newblock {\em OR Spectrum}, DOI 10.1007/s00291-009-0168-5, 2009.

\bibitem{Eng05a}
K.~Engel and E.~Tabbert.
\newblock Fast simultaneous angle, wedge, and beam intensity optimization in
  inverse radiotherapy planning.
\newblock {\em Optimization and Engineering}, 6(4):393--419, 2005.

\bibitem{Engelbeen08}
C.~Engelbeen and S.~Fiorini.
\newblock Constrained decompositions of integer matrices and their applications
  to intensity modulated radiation therapy.
\newblock {\em Networks}, DOI 10.1002/net.20324, 2009.

\bibitem{Eng_Fio_Kie09}
C.~Engelbeen, S.~Fiorini, and A.~Kiesel.
\newblock A closest vector problem arising in radiation therapy planning.
\newblock {\em Journal of Combinatorial Optimization}, DOI
  10.1007/s10878-010-9308-8, 2010.

\bibitem{Gau06}
T.~Gauer, D.~Albers, F.~Cremers, R.~Harmansa, R.~Pellegrini, and R.~Schmidt.
\newblock Design of a computer-controlled multileaf collimator for advanced
  electron radiotherapy.
\newblock {\em Phys. Med. Biol.}, 53:5987--6003, 2006.

\bibitem{Gau08}
T.~Gauer, J.~Sokoll, F.~Cremers, R.~Harmansa, M.~Luzzara, and R.~Schmidt.
\newblock Characterization of an add-on multileaf collimator for electron beam
  therapy.
\newblock {\em Phys. Med. Biol.}, 53:1071--1085, 2008.

\bibitem{Kal05}
T.~Kalinowski.
\newblock A duality based algorithm for multileaf collimator field segmentation
  with interleaf collision constraint.
\newblock {\em Discrete Appl. Math.}, 152(1-3):52--88, 2005.

\bibitem{Kal05a}
T.~Kalinowski.
\newblock Reducing the number of monitor units in multileaf collimator field
  segmentation.
\newblock {\em Phys. Med. Biol.}, 50(6):1147--1161, 2005.

\bibitem{Kal08}
T.~Kalinowski.
\newblock Multileaf collimator shape matrix decomposition. {I}n
  \emph{Optimization in Medicine and Biology}.
\newblock {\em G.J. Lim and E.K.Lee}, Auerbach Publishing:\ 253--286, 2008.

\bibitem{Kal08a}
T.~Kalinowski.
\newblock Reducing the tongue-and-groove underdosage in {MLC} shape matrix
  decomposition.
\newblock {\em Algorithmic Operations Research}, 3(2), 2008.

\bibitem{Kal04b}
T.~Kalinowski.
\newblock The complexity of minimizing the number of shape matrices subject to
  minimal beam-on time in multileaf collimator field decomposition with bounded
  fluence.
\newblock {\em Discrete Appl. Math.}, 157:2089--2104, 2009.

\bibitem{Kal08c}
T.~Kalinowski.
\newblock A min cost network flow formulation for approximated {MLC}
  segmentation.
\newblock {\em Networks}, DOI 10.1002/net.20394, 2010.

\bibitem{KalKie08}
T.~Kalinowski and A.~Kiesel.
\newblock Approximated {MLC} shape matrix decomposition with interleaf
  collision constraint.
\newblock {\em Algorithmic Operations Research}, 4(1):49--57, 2009.

\bibitem{Kam03a}
S.~Kamath, S.~Sahni, J.~Li, J.~Palta, and S.~Ranka.
\newblock Leaf sequencing algorithms for segmented multileaf collimation.
\newblock {\em Phys. Med. Biol.}, 48(3):307--324, 2003.

\bibitem{Kam04a}
S.~Kamath, S.~Sahni, J.~Palta, S.~Ranka, and J.~Li.
\newblock Optimal leaf sequencing with elimination of tongue--and--groove
  underdosage.
\newblock {\em Phys. Med. Biol.}, 49:N7--N19, 2004.

\bibitem{Kam04b}
S.~Kamath, S.~Sahni, S.~Ranka, J.~Li, and J.~Palta.
\newblock A comparison of step--and--shoot leaf sequencing algorithms that
  eliminate tongue--and--groove effects.
\newblock {\em Phys. Med. Biol.}, 49:3137--3143, 2004.

\bibitem{Lim07}
J.~Lim, M.C. Ferris, S.J. Wright, D.M. Shepard, and M.A. Earl.
\newblock An optimization framework for conformal radiation treatment planning.
\newblock {\em Informs Journal on Computing}, 19(3):366--380, 2007.

\bibitem{Mat}
M.M. Matuszak, E.W. Larsen, K.~Jee, D.L. McShan, and B.A. Fraass.
\newblock Adaptive diffusion smoothing: A diffusion based method to reduce imrt
  field complexity.
\newblock {\em Med. Phys.}, 35(4):1532--1546, 2008.

\bibitem{Nus06}
M.~Nu{\ss}baum.
\newblock Min cardinality {C}1 decomposition of integer matrices.
\newblock Master's thesis, Faculty for Mathematics, {TU} Kaiserslautern, 2006.

\bibitem{Que04}
W.~Que, J.~Kung, and J.~Dai.
\newblock `{T}ongue-and-groove' effect in intensity modulated radiotherapy with
  static multileaf collimator fields.
\newblock {\em Phys. Med. Biol.}, 49:399--405, 2004.

\bibitem{She02}
D.M. Shepard, M.A. Earl, X.A. Li, S.~Naqvi, and C.~Yu.
\newblock Direct aperture optimization: A turnkey solution for step-and-shoot
  {IMRT}.
\newblock {\em Med. Phys.}, 29(6):1007--1018, 2002.

\bibitem{Wunderling}
R.~Wunderling.
\newblock {\em Paralleler und objektorientierter Simplex-Algorithmus}.
\newblock PhD thesis, Technische Universit{\"a}t Berlin, 1996.
\newblock \url{http://www.zib.de/Publications/abstracts/TR-96-09/}.

\end{thebibliography}

\section{Appendix}

\begin{algorithm}[H]\caption{LOC-left}
\label{alg:loc}
\begin{algorithmic}
\REQUIRE vector $\bm{v}$
\FOR {$i=min$ to $max$}
   \STATE $tc_{i1}=|v_i-i|$
\ENDFOR
\STATE $tc_{ij}=\infty$ for all $i \in [min,max],j>1$
\FOR {$j=2$ to $k$}
   \FOR {$i=min$ to $max$}
      \FOR {$i'=min$ to $i$}
         \IF {$tc_{i,j-1}+|v_j-i'|<tc_{i'j}$}
            \STATE $tc_{i'j}=tc_{i,j-1}+|v_j-i'|$
            \STATE $pre_{i'j}=i$
         \ENDIF
      \ENDFOR
   \ENDFOR
\ENDFOR
\STATE $opt=\min_{i \in [min,max]} tc_{i,k}$
\STATE Let $i_{opt}$ be one of the indices with $tc_{i_{opt},k}=opt$.
\FOR {$j=k$ down to $1$}
   \STATE $w_j=i_{opt}$
   \STATE If still $j>1$, then $i_{opt}=pre_{i_{opt},j}$
\ENDFOR
\ENSURE vector $\bm{w}$
\end{algorithmic}
\end{algorithm}

\begin{algorithm}[H]\caption{Segmentation ASS}
\label{alg:seg_ASS}
\begin{algorithmic}
\REQUIRE matrix $B$
\WHILE {$B \neq 0$}
   \STATE $i=1;$
   \WHILE {$\bm{b}_i=\bm{0}$}
      \STATE $l_i=n+1$; $r_i=0$; $i=i+1$;
   \ENDWHILE
   \STATE Choose $l_i$ and $r_i$ for row $i$ such that $r_i-l_i \ge g_1-1$ (see Remark 2).
   \WHILE {$r_i-l_i \ge g_1-1$ and $\bm{b_{i+1}} \neq \bm{0}$}
      \STATE {\bf Find interval ASS}: Choose $l_{i+1}$ and $r_{i+1}$ for row $i+1$ such that the overlap with the previous row is $\ge g_1$ or close row $i+1$.
      \STATE $i=i+1$;
   \ENDWHILE
   \STATE Close all remaining rows with $l_i=n+1$ and $r_i=0$.
   \FOR {$i=1$ to $m$}
      \FOR {$j=l_i$ to $r_i$}
         \STATE $b_{ij}=b_{ij}-1$;
      \ENDFOR
   \ENDFOR
   \STATE Store $((l_1,r_1),\dots,(l_m,r_m))$ in the segmentation.
\ENDWHILE
\ENSURE Segmentation
\end{algorithmic}
\end{algorithm}

\begin{algorithm}[H]\caption{Find interval ASS}
\label{alg:find_interval_ASS}
\begin{algorithmic}
\REQUIRE $i, l_i, r_i$
\STATE $finish=0;$
\WHILE {$finish=0$}
     \STATE $start=\min \{ j \ | \ b_{i+1,j}>0 \}$;
     \STATE $end=\min \{ j \ | \ b_{i+1,j}>b_{i+1,j+1} \}$;
     \STATE $l=\max(start,l_i)$;
     \STATE $r=\min(end,r_i)$;
     \IF {$r-l \ge g_1-1$}
        \STATE $l_{i+1}=start$; $r_{i+1}=end$; $finish=1$;
     \ELSIF {$r-l<0$}
        \STATE $l_{i+1}=n+1$; $r_{i+1}=0$; $finish=1$;
     \ELSE
        \IF {$start>l_i$}
           \STATE $b_{i+1,start-1}=b_{i+1,start-1}+1$;
        \ELSE
           \STATE $b_{i+1,end+1}=b_{i+1,end+1}+1$;
        \ENDIF
     \ENDIF
\ENDWHILE
\ENSURE $l_{i+1}$, $r_{i+1}$
\end{algorithmic}
\end{algorithm}

\vspace{1em}
{\bf Remark 2.} The interval $[l,r]$ for the first open row is computed analogously to the interval for the other rows, only ignoring the overlap constraint and instead requiring $r-l>g_1-1$.
\vspace{1em}

\begin{algorithm}[H]\caption{Segmentation ASAS}
\label{alg:seg_ASAS}
\begin{algorithmic}
\REQUIRE matrix $B$
\WHILE {$B \neq 0$}
   \STATE $i=1$; $s_{ij}=0$ for all $i$ and $j$;
   \WHILE {$\bm{b_i}=\bm{0}$}
      \STATE $l_i=n+1$; $r_i=0$; $i=i+1$;
   \ENDWHILE
   \STATE $start=i$; $closed=0;$
   \IF {$start > n-g_2+1$}
      \STATE $closed=1$;
   \ELSE
      \STATE Choose $l_i$ and $r_i$ for row $i$ such that $r_i-l_i \ge g_1-1$ or close row $start$ (see Remark 3).
      \IF {row $start$ is closed}
         \STATE $closed=1$;
      \ELSE
         \WHILE {$r_i-l_i \ge g_1-1$ and $\bm{b_{i+1}} \neq \bm{0}$}
            \STATE {\bf Find interval ASAS}: Choose $l_{i+1}$ and $r_{i+1}$ for row $i+1$ such that the overlap with
                                             the previous row is larger than $g_1$ or close row $i+1$.\\
            \STATE $i=i+1$;
         \ENDWHILE
         \FOR {all remaining rows $i$}
            \IF {$\bm{s_i} \neq \bm{0}$}
               \STATE $l_i=$index of the first one in $\bm{s_i}$; $r_i=$index of the last one in $\bm{s_i}$;
            \ELSE
               \STATE $l_i=n+1$, $r_i=0$;
            \ENDIF
         \ENDFOR
      \ENDIF
   \ENDIF
   \IF {$closed=1$}
      \STATE $\bm{b_{start}}=\bm{0}$;
   \ELSE
      \FOR {$i=1$ to $m$}
         \FOR {$j=l_i$ to $r_i$}
            \IF {$b_{ij}>0$}
               \STATE $b_{ij}=b_{ij}-1$;
            \ENDIF
         \ENDFOR
      \ENDFOR
      \STATE Store $((l_1,r_1),\dots,(l_m,r_m))$ in the segmentation.
   \ENDIF
\ENDWHILE
\ENSURE Segmentation
\end{algorithmic}
\end{algorithm}

\begin{algorithm}[H]\caption{Find interval ASAS}
\label{alg:find_interval_ASAS}
\begin{algorithmic}
\REQUIRE $i+1, l_i, r_i$, current matrix $S$
\IF {$\bm{s_i}=\bm{0}$}
   \STATE $v_1=n+1$; $v_2=0$;
\ELSE
   \STATE $v_1=$ index of the first one in $\bm{s_i}$; $v_2=$ index of the last one in $\bm{s_i}$;
\ENDIF
\STATE $t_1=$ index of the first positive entry in $\bm{b_i}$;
\STATE $t_2=$ index of the last positive entry in $\bm{b_i}$;
\STATE $opt=0$; $l_{i+1}=n+1$; $r_{i+1}=0$;
\FOR {$l=\min(v_1,t_1)$ to $\min(b_l,v_1)$}
   \FOR {$r=\max(l+g_1-1,v_2,b_r)$ to $\max(v_2,t_2)$}
      \IF {$\min(r,r_i)-\max(l,l_i) \ge g_1-1$ and not ($l<l_1$ and $i+1>m-g_2+1$) and not ($r>r_i$ and $i+1>m-g_2+1$) and no unavoidable zero is between $l$ and $r$}
         \STATE $benchmark=0$;
         \FOR {all entries $(k,j)$ of unavoidable ones with $i+2 \le k \le i+g_2$}
            \IF {$b_{kj}>0$}
               \STATE $benchmark=benchmark-1$;
            \ELSE
               \STATE $benchmark=benchmark+1$;
            \ENDIF
         \ENDFOR
         \FOR {$j=l$ to $r$}
            \IF {$b_{i+1,j}>0$}
               \STATE $benchmark=benchmark-1$;
            \ELSE
               \STATE $benchmark=benchmark+1$;
            \ENDIF
         \ENDFOR
         \IF {$benchmark<opt$}
            \STATE $opt=benchmark$; $l_{i+1}=l$; $r_{i+1}=r$;
         \ENDIF
      \ENDIF
   \ENDFOR
\ENDFOR
\STATE Put unavoidable ones or zeros corresponding to the decision for $l_{i+1}$ and $r_{i+1}$ into $S$.
\ENSURE $l_{i+1}$, $r_{i+1}$, current matrix $S$
\end{algorithmic}
\end{algorithm}

\vspace{1em}
{\bf Remark 3.} The choice of $(l,r)$ in row $start$ is again computed analogously, only ignoring the overlap constraint with the previous row.
\vspace{1em}

\begin{algorithm}[H]\caption{Total Change Improvement}
\label{alg:improvement}
\begin{algorithmic}
\REQUIRE Segmentation
\FOR {all segments}
   \FOR {$i=1$ to $m$}
         \IF {$l_i>1$ and $b_{i,l_i-1}<a_{i,l_i-1}$\\ \quad(for ASAS: and decreasing $l_i$ will not violate (iv))}
            \STATE $l_i=l_i-1;$
         \ENDIF
         \IF {$r_i<1$ and $b_{i,r_i+1}<a_{i,r_i+1}$\\ \quad (for ASAS: and increasing $r_i$ will not violate (iv))}
            \STATE $r_i=r_i+1;$
         \ENDIF
         \IF {$b_{i,l_i}>a_{i,l_i}$ and increasing $l_i$ will not violate (iii)\\ \quad (for ASAS: and increasing $l_i$ will not violate (iv))}
            \STATE $l_i=l_i+1;$
         \ENDIF
         \IF {$b_{i,r_i}>a_{i,r_i}$ and decreasing $r_i$ will not violate (iii)\\ \quad (for ASAS: and decreasing $r_i$ will not violate (iv))}
            \STATE $r_i=r_i-1;$
         \ENDIF
   \ENDFOR
   \STATE Update approximation matrix
\ENDFOR
\ENSURE Segmentation
\end{algorithmic}
\end{algorithm}

\end{document}